\documentclass[11pt]{article}
\usepackage{epsfig,amssymb,graphicx,amsmath,amsthm,subcaption}
\usepackage{tikz}
\usepackage{float}
\usetikzlibrary{shapes.geometric, arrows}
\usetikzlibrary{positioning}
\usepackage{enumerate}
\usepackage{calrsfs}

\newtheorem{theorem}{Theorem}[section]

\newtheorem*{remark}{Remark}
\usepackage[mathcal]{eucal}

\begin{document}
		\title{Time-delayed model of RNA interference}
		
\author{G. Neofytou,\hspace{0.5cm}Y.N. Kyrychko,\hspace{0.5cm}K.B. Blyuss\thanks{Corresponding author. Email: k.blyuss@sussex.ac.uk} 
\\\\ Department of Mathematics, University of Sussex, Falmer,\\
Brighton, BN1 9QH, United Kingdom}

\maketitle

	\begin{abstract}
		RNA interference (RNAi) is a fundamental cellular process that inhibits gene expression through
		cleavage and destruction of target mRNA. It is responsible for a number of important intracellular functions,
		from being the first line of  immune defence against pathogens to regulating development and morphogenesis.
		In this paper we consider a mathematical model of RNAi with particular emphasis on time delays associated
		with two aspects of primed amplification: binding of siRNA to aberrant RNA, and binding of siRNA to mRNA,
		both of which result in the expanded production of dsRNA responsible for RNA silencing. Analytical and numerical
		stability analyses are performed to identify regions of stability of different steady states and to determine
		conditions on parameters that lead to instability. Our results suggest that while the original model without time
		delays exhibits a bi-stability due to the presence of a hysteresis loop, under the influence of time delays, one of the two
		steady states with the high (default) or small (silenced) concentration of mRNA can actually lose its stability via a Hopf bifurcation.
		This leads to the co-existence of a stable steady state and a stable periodic orbit, which has a profound effect on the
		dynamics of the system.
	\end{abstract}
	
	\section{Introduction}
	
	RNA interference is a complex biological process that occurs in  many eukaryotes and fulfils a regulatory role by allowing control over gene expression \cite{He2004,Hannon2002,Ketting2001}, while also providing an effective immune response against viruses and tranposons through its ability to target and destroy specific mRNA molecules \cite{Mandadi2013,Agius2012}. This multi-step process is mediated by double-stranded RNAs (dsRNA) of different lengths that are generated by an inverted-repeat transgene, or an invading virus during its replication process \cite{Vaucheret2001,Sharma2013}. A very simple description of the core pathway is as follows. The presence of transgenic or viral dsRNA triggers an immune response within the host cell, whereby the foreign RNA is targeted by specialized enzymes called dicers (DLC). These enzymes cleave the target RNA into short 21-26 nucleotide long molecules, named short interfering RNAs (siRNA) or microRNA (miRNA), which can subsequently be used to assemble a protein complex, called the RNA-induced silencing complex (RISC). This specialized complex can recognise and degrade RNAs containing complementary sequences into garbage RNA that can no longer be translated into a functioning protein, thus leading to the translational arrest of the viral or transgenic RNA \cite{Escobar2000,Elbashir2001}. While the core pathway might be sufficient to describe RNA interference in mammals, for other organisms it is possible that the process is not strictly limited to the molar concentration of siRNA at the initiating site, but can spread systemically \cite{Palauqui1997,Melnyk2011,Zhang2012}.
	
	 In the studies of RNA interference in the nematode {\it Caenorhabditis elegans}, it was observed that a notable portion of the produced siRNA was not derived directly from the initializing dsRNA, suggesting a presence of a mechanism in which some additional dsRNA could be generated \cite{Sijen2001}. To account for this discovery,  primed and unprimed amplification pathways were proposed, in which an RNA-dependent RNA polymerase (RdRp) or RNA replicase could synthesize the additional unaccounted dsRNA \cite{Lipardi2001,Makeyev2002}. In the case of primed amplification, it is postulated that when assisted by RdRP, the siRNA which binds on mRNA can itself initialise dsRNA synthesis, thus generating a new round of dsRNAs ready to be used in the process. On the other hand, unprimed amplification describes the situation where dsRNA synthesis occurs without the assistance of the primer RdRP, but instead relies  on the presence of garbage RNA to facilitate synthesis. As in most complex biological processes, RNAi carries risks and is prone to different errors, as it necessitates the host's ability to correctly discriminate between endogenous and exogenous mRNA \cite{Giordano2002}. Thus, any invading viral sequences with cross-reactive similarities  or  accidental  production of anti-sense transcripts corresponding to self genes can result in a self-reactive response that can be extremely damaging to the host. To limit the self-damage caused by the feed-forward amplification in RNAi, a protection mechanism has been proposed in \cite{Pak2012}. 
	 
	 A number of mathematical models have considered different aspects of RNAi in its roles of immune guard against viral infections, as well as an attractive tool for targeted gene silencing that is important for gene therapies. One of the earliest models was developed and analysed by Bergstrom et al. \cite{Bergstrom2003}. These authors focused on the issue of avoiding self-directed gene silencing during RNAi and hypothesised that this can be achieved via {\it unidirectional amplification}, whereby silencing only persists in the presence of a continuing input of dsRNA, thus acting as a safeguard against a sustained self-damaging reaction, or, in the case of viral infection, ending the process once the infection is cleared. This model was extended by Groenenboom et al. \cite{Groenenboom2005}, who analysed primed and unprimed amplification pathways to account for the dsRNA dosage-dependence of RNAi and to correctly describe the nature of transient and sustained silencing. Groenenboom and Hogeweg \cite{Groen08b} and Rodrigo et al. \cite{rodrigo} have analysed how viral replication is affected by its interactions with RNAi for plus-stranded RNA viruses, with particular account for different viral strategies for evading host immune response.
	 
	 Similarly to natural or artificial control systems, biological systems also possess intrinsic delays that arise from the lags in the sensory process of response-initiating variables, the transportation of components that regulate biological interactions, after-effect phenomena in inner dynamics and metabolic functions, including the times necessary for synthesis, maturation and reproduction of cells and whole organisms \cite{Ric03,Jus10,PBK15}. These delays can often lead to changes in stability and play a significant role in modelling control systems that typically involve a feedback loop. On the other hand, mathematical models without time delays are based on the assumption that the transmission of signals and biological processes occur instantaneously. Although the timescale associated with these delays can sometimes be ignored, for instance, when the characteristic timescales of the model are very large compared to the observed delays, there are clear cases where the present and future state of a system depend on its past history. In such situations, dynamics of the system can only be accurately described with delay differential equations rather than the traditional ordinary differential equations. Due to the non-instantaneous nature of the complex processes involved in RNA interference, it is biologically feasible to explicitly include time delays associated with the times required for transport of RNAi components, and assembly of different complexes. Nikolov and Petrov \cite{Nikolov2007} and Nikolov et al. \cite{Nik09} have considered the effects of such time delays within a single amplification pathway as modelled by Bergstrom et al. \cite{Bergstrom2003}. Under a restrictive and somewhat unrealistic assumption that the natural degradation of RISC-mRNA complex takes place at exactly the same speed as formation of new dsRNA, the authors have shown how time delays can induce instability of the model steady state, thus disrupting gene silencing and causing oscillations.
	 
	 In the context of siRNA-based treatment, Bartlett and Davis \cite{bartlett} have performed a detailed analysis of the process of siRNA delivery and its interaction with the RNAi machinery in mammalian cells, and compared it to experimental results in mural cell cultures. This model and associated experiments have provided significant insights into optimising the dosage and scheduling of the therapeutic siRNA-mediated gene silencing. Raab and Stephanopoulos \cite{raab} also considered siRNA dynamics in mammalian cells with an emphasis on two-gene systems with different kinetics for the two genes. Arciero et al. \cite{arciero} studied a model of siRNA-based tumour treatment which targets the expression of TGF-$\beta$, thus reducing tumour growth and enhancing immune response against tumour cells.
	 
	 Since originally RNA interference was discovered in plants \cite{napoli}, which present a very convenient framework for experimental studies of RNAi, a number of mathematical models have considered specific aspects of the dynamics of viral growth and its interactions with RNAi in plants. Groenenboom and Hogeweg \cite{Groen08a} have analysed a detailed model for the dynamics of intra- and inter-cellular RNA silencing and viral growth in plants. This spatial model has demonstrated different kinds of infection patterns that can occur on plant leaves during viral infections. More recently, Neofytou et al. \cite{NKB16a} have analysed the effects of time delays associated with the growth of new plant tissue and with the propagation of the gene silencing signal. They have shown that a faster propagating silencing signal can help the plant recover faster, but by itself is not sufficient for clearance of infection. On the other hand, a slower silencing signal can lead to sustained periodic oscillations around a chronic infection state. In a very important practical context of viral co-infection, Neofytou et al. \cite{NKB16b} have studied how the dynamics of two viruses simultaneously infecting a single host is mediated by the RNAi.
	 
	In this paper we consider a model of RNAi with primed amplification, and focus on the role of two time delays associated with the production of dsRNA directly from mRNA, or from aberrant RNA. An important result obtained in this study is partial destruction of the hysteresis loop: while the original model without time delays is bi-stable, under the influence of time delays, the steady state with either the smallest or the highest concentration of mRNA can lose its stability via a Hopf bifurcation. This leads to the co-existence of a stable steady state and a stable periodic orbit, which has a profound effect on the dynamics of the system. When the default steady state is destabilized by the time delays, our numerical analysis shows that the system will always converge to the silenced steady state. On the other hand, in parameter regimes where time delays destabilize the silenced steady state, the system will either converge to the default steady state, or it will oscillate around the unstable steady state depending on the initial conditions. In fact, under the influence of time delays, one would requires an even higher initial dosage of dsRNA to achieve sustained silencing. However, when there is stable periodic orbit around the silenced steady state, one would also have to consider the amplitude of these oscillations and how it may affect the phenotypic stability of the species in question. Thus, the augmented model exhibits an enriched dynamical behavior compared to its predecessor which otherwise can only be replicated by different extensions to the core pathway, like the RNase model developed in \cite{Groenenboom2005}, which assumes the presence of a specific siRNA-degrading RNase with saturating kinetics. The outline of the paper is as follows. In the next section we introduce the model and discuss its basic properties. In Section 3 we identify all steady states of the model together with conditions for their biological feasibility. Sections 4 and 5 are devoted to the stability analysis of these steady states depending on model parameters, including numerical bifurcation analysis and simulations of the model that illustrate different types of dynamical behaviour. The paper concludes in Section 6 with the discussion of results and open problems.

\section{Model derivation}

To analyse the dynamics of RNAi with primed amplification, following Groenenboom et al. \cite{Groenenboom2005} we consider the populations of mRNA, dsRNA, siRNA and garbage (aberrant) RNA, to be denoted by $M(t)$, $D(t)$, $S(t)$ and $G(t)$, respectively. It is assumed that mRNA is constantly transcribed by each transgene at rate $h$, with $n_1$ being the number of transgenic copies, and is degraded at the rate $d_m$. For simplicity, it will be assumed that each transgene produces the same amount of mRNA. Some dsRNA is synthesized directly from mRNA through the activity of RdRp at a rate $p$. The available dsRNA is cleaved by a dicer enzyme into $n_2$ siRNA molecules at a rate $a$.  In this model it is assumed that siRNA is involved into forming two distinct complexes that use the siRNA as a guide to identify and associate with different categories of RNA strands to initiate the dsRNA synthesis. The first is the RISC complex responsible for degrading mRNA into garbage RNA, which decays naturally at a rate $d_g>d_m$. For simplicity, the RISC population is not explicitly included in the model, but it is rather assumed that siRNA directly associates with mRNA at a rate $b_1$. The second complex guided by  siRNA binds mRNA aberrant (garbage) RNA, and subsequently is primed by RdRp to synthesize additional dsRNA (primed amplification). To avoid unnecessary complexity, the second complex will also be represented implicitly by assuming that siRNA directly associates with mRNA and garbage RNA for the purpose of dsRNA synthesis at the rates $b_2$ and $b_3$ respectively. At this point, we include two distinct time delays $\tau_1$ and $\tau_2$ to represent the delays inherent in the the production of dsRNA from mRNA and garbage RNA, respectively. With these assumptions, the system describing the dynamics of different RNA populations takes the form
\begin{equation}\label{system:garbage}
\begin{array}{l}
\displaystyle{\frac{dM}{dt} =n_1h- d_m M(t) - pM(t) - b_1S(t)M(t) - b_2S(t)M(t),}\\\\
\displaystyle{\frac{dD}{dt}  = pM(t) - aD(t) + b_2S(t-\tau_1)M(t-\tau_1) + b_3S(t-\tau_2)G(t-\tau_2),}\\\\
\displaystyle{\frac{dS}{dt}  = n_2aD(t) - d_sS(t) -  b_1S(t)M(t) - b_2S(t)M(t)- b_3S(t)G(t),}\\\\
\displaystyle{\frac{dG}{dt}  = n_3b_1S(t)M(t) - d_gG(t) - b_3S(t)G(t),}
\end{array}
\end{equation}
with the initial conditions
\begin{equation}\label{IConds}
\begin{array}{l}
M(s) =M_0(s)\ge0,\quad s\in[-\tau_1,0],\quad G(s) = G_0(s)\ge 0,\quad s\in[-\tau_2,0],\\\\
S(s)=S_0(s)\ge 0,\quad s\in[-\tau,0],\quad \tau = \max\{\tau_1,\tau_2\},\quad D(0)\ge0.
\end{array}
\end{equation}
Before proceeding with the analysis of the model (\ref{system:garbage}), we have to establish that this system is well-posed, i.e. its solutions are non-negative and bounded.

\begin{remark}
Invariance of the positive orthant follows straightforwardly from the Theorem 5.2.1 in \cite{Smith1995}.
Existence, uniqueness and regularity of solutions $M(t),\;D(t),\; S(t),\; G(t)$ of the system (\ref{system:garbage}) with the initial conditions (\ref{IConds}) follow from the standard theory discussed in \cite{Kuang93,Smith10}.
\end{remark}

\begin{theorem}\label{theo:boundedness}
Suppose there exists a time $T>0$, such that the solution $D(t)$ of the model (\ref{system:garbage}) satisfies the condition $D(t)\le \widehat{D}$ for all $t\ge T$ with $\widehat{D}>0$. Then, the solutions $M(t),\; S(t),\; G(t)$ of the model (\ref{system:garbage}) are bounded for all $t\ge T$.
\end{theorem}

\noindent {\bf Proof.} Suppose $t\ge T$. Using the non-negativity of solutions, one can rewrite the first equation of the system (\ref{system:garbage}) in the form
	\[
	\displaystyle{\frac{dM}{dt}  \le n_1 h-(d_m+p)M(t)\quad\Longrightarrow\quad M(t)\leq \widehat{M}=\frac{n_1 h}{d_m+p}+M(0),}
	\]
	which shows that $M(t)$ is also bounded for $t\ge T$. The last equation of (\ref{system:garbage}) can now be rewritten as follows
	\[
	\displaystyle{\frac{dG}{dt} \le S(t)\left[n_3 b_1\widehat{M}-b_3 G(t)\right]-d_g G(t).}
	\]
	Since $S(t)\geq 0$, this inequality suggests that if $G(0)<\widehat{G}=n_3b_1\widehat{M}/b_3$, then initially it may increase, but it will never reach the value of $\widehat{G}$. Similarly, if initially $G(0)\geq \widehat{G}$, then $G$ would be monotonically decreasing, and once its value is below $\widehat{G}$, it would never go above it. Hence, $G$ is also bounded for $t\ge T$.

The third equation of the system (\ref{system:garbage}) can be recast in the form
	\[
	\frac{dS}{dt}  \le n_2a\widehat{D} - d_sS(t).
	\]
	Using the assumption of boundedness of $D$ and the comparison theorem, one then has
	\[
	\displaystyle{S(t)\le \frac{n_2a\widehat{D}}{d_s}\left(1-e^{-d_st}\right) + S(0)e^{-d_st}\le \widehat{S}=\frac{n_2a\widehat{D}}{d_s} + S(0),}
	\]
	which implies that $S(t)$ is bounded for $t\ge T$. 
	
	Hence, one concludes the existence of upper bounds $\widehat{S}$, $\widehat{M}$ and $\widehat{G}$, such that $S(t)\le \widehat{S}$, $M(t)\le \widehat{M}$ and $G(t)\le \widehat{G}$ for all $t\ge T$, which concludes the proof.
	
\hfill$\blacksquare$

\begin{remark}
In all our numerical simulations, including the ones presented in
Section 5, the solutions of the system (\ref{system:garbage}) always satisfied the condition that $D(t)$ remains bounded, which, in light of {\bf Theorem \ref{theo:boundedness}}, implies boundedness of all other state variables.
\end{remark}

\section{Steady states and their feasibility}

Steady states of the system (\ref{system:garbage}) are given by non-negative roots of the following system of algebraic equations
\begin{equation}\label{SSeq}
\begin{array}{l}
\displaystyle{n_1h- d_m M- pM- b_1SM - b_2SM=0,}\\\\
\displaystyle{pM - aD + b_2SM+ b_3SG=0,}\\\\
\displaystyle{n_2aD - d_sS -  b_1SM - b_2SM- b_3SG=0,}\\\\
\displaystyle{n_3b_1SM - d_gG - b_3SG=0.}
\end{array}
\end{equation}
It is straightforward to see that the system (\ref{SSeq}) does not admit solutions with $M=0$, as this would immediately violate the first equation due to the presence of the constant transcription of mRNA. Substituting $S = 0$ into the third equation implies $D=0$, and due to the second equation this then implies $M=0$, which is impossible. Hence, there can be no steady states with either $D$ or $S$ being zero. Similarly, if $G = 0$, the last equation implies $SM = 0$ which again is not possible. Thus, the system can only exhibit steady states where all components are non-zero.

Let us introduce the following auxiliary parameters
\begin{equation}
b = b_1 + b_2, \qquad \hat{h} = n_1h.
\end{equation}
Assuming  $S^*\ge0$, one can solve the first equation of (\ref{SSeq}) to obtain 
\begin{equation}\label{formu:M}
M^*=M(S^*)=\frac{\hat{h}}{p + d_m + bS^*}>0.
\end{equation}
Adding the second and the third equations of the system (\ref{SSeq}) gives 
\begin{equation}\label{formu:D}
D^* =D(S^*)=\frac{bd_s{S^*}^2 + (\hat{h}b_1 + pd_s + d_md_s)S^* - ph}{a[p+d_m +bS^*](n_2 - 1)}.
\end{equation}
One should note that for $S^*\ge0$ and $n_2\ge1$, $D^*\ge0$ if and only if the following condition holds
 \begin{equation}
 bd_s{S^*}^2 + (b_1h\hat{h} + d_md_s + d_sp)_{z}S^* - hp>0, 
 \end{equation}
 which implies that $S^*$ must satisfy
 \begin{eqnarray}
 S^*\ge S_{\rm min}={\frac {-z+\sqrt {4\,bd_{{s}}ph+{z}^{2}}}{2bd_{{s}}}},\; z =b_1h\hat{h} + d_md_s + d_sp. 
 \end{eqnarray}
From the last equation of the system (\ref{SSeq}) and using the expression for $M^*$ we obtain
\begin{gather}\label{formu:G}
\begin{aligned}
G^*=G(S^*)= \frac{\hat{h}n_3b_1S^*}{(p+d_m +bS^*)(b_3S^* + d_g)}>0. 
    \end{aligned}
\end{gather}
Substituting these values back into the third equation of the system (\ref{SSeq}) one obtains the following cubic equation for $S^*$
\begin{equation}\label{poly:Q(S)}
Q(S^*)= \alpha_3S^3 + \alpha_2 S^2 + \alpha_1S + \alpha_0=0,
\end{equation}
where
\[
\begin{array}{l}
\alpha_0= -\hat{h}pd_gn_2<0,\quad \alpha_3 = bb_3d_s>0,\\
\alpha_1=\hat{h}[d_gb - n_2(pb_3 + d_gb_2)] + d_gd_s(p + d_m),\\
\alpha_2=\hat{h}[b_1b_3(1+n_3-n_2n_3) + b_2b_3(1-n_2)] +b_3d_s(p+ d_m) + bd_gd_s.
\end{array}
\]
It is obvious that the cubic $Q(S^*)$ has at least one positive real root for any $n_i\ge1, \; i =1,2,3.$ In fact, by using Descartes's  rule of signs one can deduce that this cubic has exactly one positive and two negative roots, with the exception of $\alpha_2<0$ and $\alpha_1>0$, when it admits three positive roots. We can summarise this in the following theorem.
\begin{theorem}\label{Theorem:Steady states}
Let \begin{equation}
\Delta = 18\alpha_3\alpha_2\alpha_1\alpha_0 -4\alpha_2^3\alpha_0 + \alpha_2^2\alpha_1^2 - 4\alpha_3\alpha_1^3 - 27\alpha_3^2\alpha_0^2
\end{equation}
be the discriminant of equation (\ref{poly:Q(S)}). Then equation (\ref{poly:Q(S)}) has three distinct real roots if and only if $\Delta \ge 0 $, and it has three real roots with one double root if $\Delta = 0$. Therefore, there will be a single feasible equilibrium if either $\alpha_2\ge 0 $; or $\alpha_1\le 0$; or $\alpha_2 <0$, $\alpha_1>0$, and $\Delta<0$. On the other hand, if $\alpha_2<0$ and $\alpha_1>0$, and $\Delta>0$, then there are exactly three distinct feasible equilibria. For the degenerate situation of $\Delta = 0$, when $\alpha_2<0$ and $\alpha_1>0$, anything between one and three distinct feasible equilibria is possible.
\end{theorem}
\section{Stability  analysis}

Linearisation of the delayed system (\ref{system:garbage}) around the steady state $E=(M^*,D^*,S^*,R^*)$ yields the following characteristic equation\begin{equation}\label{sys:nodelays}
P(\lambda) = p_4\lambda^4 + p_3\lambda^3 + p_2\lambda^2 + p_1\lambda +p_0=0,
\end{equation}
where the coefficients $p_i$, $i = 0,..,4$, are given in the Appendix, and for convenience of notation we have dropped stars next to the steady state values and introduced auxiliary parameters $T_i = e^{-\lambda\tau_i}$, $i=1,2$. In the case of instantaneous primed amplification, i.e. for $T_{1,2} = 1$  in (\ref{sys:nodelays}), any  steady state $(M^*,S^*,D^*,G^*)$ defined in {\bf Theorem~\ref{Theorem:Steady states}} is linearly asymptotically stable, if the appropriate Routh-Hurwitz conditions are satisfied, i.e if $p_0,\ldots, p_4>0$, $p_3p_2>p_1p_4$, and $p_3p_2p_1>p_4p_1^2 + p_3^2p_0$.

\subsection{Single primed amplification delay}

As a first case, we consider a situation where one of the primed amplification delays is negligibly small compared to other timescales of the model, so that that part of the amplification pathway can be considered to take place instantaneously. Formally, this can be represented by $\tau_n>0$ for some $n=1,2$, with $\tau_m = 0$ for $m\ne n$. In this case, analysis of the distribution of roots of the characteristic equation follows the methodology of \cite{Ruan2001}. The first step is to rewrite the characteristic equation (\ref{sys:nodelays}) in the form
\begin{equation}
\lambda^4 + \alpha\lambda^3 + (\beta_1T_1 + \beta_2T_2 +\beta_3)\lambda^2 + (\gamma_1T_1 + \gamma_2T_2 + \gamma_3)\lambda 
+(\delta_1T_1 + \delta_2T_2 + \delta_3)=0,\\
\end{equation}
where
\[
\begin{array}{l}
\alpha = b_1n_3MSG^{-1} + an_2DS^{-1} + a + \hat{h}M^{-1},\\\\
\beta_1 = -ab_2n_2M, \qquad \beta_2 =-an_2b_3G,\\\\
\displaystyle{\beta_3 = \frac{ab_1n_3(n_2D + S)M}{G}\! +\! \frac{an_2(aM + \hat{h})D}{MS}\!+\! \frac{\hat{h}(b_1n_3SM + aG)}{MG}\! -\! b^2SM\! +\! b_3d_gG,}\\\\
\gamma_1 = ab_2n_2[(bG-b_1n_3)SMG^{-1} - \hat{h}], \gamma_2 = -an_2b_3[b_1n_3SM + (d_g + \hat{h}M^{-1})G],\\\\
\gamma_3 =  b b_1 n_3 M S^2(b_3 + bMG^{-1}) + an_2D[b_1n_3G^{-1}(a + \hat{h})  + a\hat{h}G^{-1}M^{-1}]\\
\hspace{1cm}+\hat{h}(ab_1n_3SG^{-1} + b_3d_g GM^{-1}) + a[b_3d_gG + M(b^2S + bpn_2)],\\\\
\delta_1 = ab_1b_2n_2n_3MS(bSM - \hat{h})G^{-1},\quad \delta_2 = an_2b_3[b_1n_3S(bMS -\hat{h}) - \hat{h}d_gGM^{-1}],\\\\
\delta_3 = abb_1n_3MS[pn_2MG^{-1} - S(b_3 + bMG^{-1})] + a\hat{h}(ab_1n_2n_3G^{-1} + b_3d_g GM^{-1}).
\end{array}
\]
If one of the delays $\tau_m$ is zero, we have
\begin{equation}\label{eq:chp_single_delay}
\lambda^4 + \alpha\lambda^3 + (\beta_nT_n + \hat{\beta}_m)\lambda^2 + (\gamma_nT_n + \hat{\gamma}_m)\lambda + (\delta_nT_n + \hat{\delta}_m)=0,
\end{equation}
where
\[
\hat{\beta}_m = \beta_m +\beta_3,\;\; \hat{\gamma}_m = \gamma_m +\gamma_3,\;\; \hat{\delta}_m = \delta_m +\delta_3.
\]
To investigate whether this equation can have purely imaginary roots, we substitute $\lambda = i\omega$ with some $\omega>0$ and separate real and imaginary parts, which yields the following system of equations
\begin{gather}\label{sys:trigonometric}
\begin{aligned}
& \omega\gamma_n \sin(\omega\tau_n) +(\delta_n-\omega^2\beta_n)\cos(\omega\tau_n) = \omega^2(\hat{\beta}_m - \omega^2)-\hat{\delta}_m,\\
& \omega\gamma_n \cos(\omega\tau_n) -(\delta_n-\omega^2\beta_n)\sin(\omega\tau_n) = \omega(\alpha \omega^2 -\hat{\gamma}_m).
\end{aligned}
\end{gather}
Squaring and adding these two equations gives the equation for the Hopf frequency $\omega$
\begin{equation}\label{equation:h(v)}
h(v)=v^4 + c_3 v^3 + c_2v^2 + c_1 v + c_0 = 0,\qquad v=\omega^2,
\end{equation}
with
\[
\begin{array}{l}
c_0 = {\hat{\delta}_m}^2-\delta_n^2,\quad c_1 = 2(\beta_n\delta_n - \hat{\beta}_m\hat{\delta}_m) +\hat{\gamma}_m^2 - \gamma_n^2,\\\\
c_2 = 2(\hat{\delta}_m - \alpha\hat{\gamma}_m) + \hat{\beta}_m^2 - \beta_n^2,\quad c_3 =\alpha^2-2\hat{\beta}_m.
\end{array}
\]
Let us assume that the equation (\ref{equation:h(v)}) has four distinct positive roots denoted by $v_1$, $v_2$, $v_3$ and $v_4$. This implies that the equation (\ref{eq:chp_single_delay}) in turn has  four purely imaginary roots $\lambda = i\omega_k$, $k=1,...,4$, where
\begin{equation}
\omega_1 = \sqrt{v_1},\qquad \omega_2 = \sqrt{v_2},\qquad \omega_3 = \sqrt{v_3},\qquad \omega_4 = \sqrt{v_4}.
\end{equation}
With the help of auxiliary parameters
\[
F_1 = w\omega\gamma_n, F_2 = \delta_n - w^2\beta_n, H_1 =  w^2(\hat{\beta}_m - w^2) - \hat{\delta}_m, H_2 = w(aw^2-\hat{\gamma}_m),
\]
one can rewrite the system (\ref{sys:trigonometric}) in the form
\begin{gather}
\begin{aligned}
F_1\sin(w\tau_n) + F_2\cos(w\tau_n) = H_1,\\
F_1\cos(w\tau_n) - F_2\sin(w\tau_n) = H_2.\\
\end{aligned}
\end{gather}
From this system we obtain 
\begin{equation}
\tan(w\tau_n) = \frac{F_1H_1 - F_2H_2}{H_1F_2 + H_2F_1},
\end{equation}
which gives the values of the critical time $\tau_n$ for each $k=1,...,4$, and any $\xi\in\mathbb{N}$ as
\begin{equation}\label{eq:critical_delay}
\begin{array}{l}
\displaystyle{\tau_{n,k}^{(\xi)} =\frac{1}{\omega_k}\Bigg[(\xi-1)\pi+}\\\\
\displaystyle{\arctan\left(\frac{(\alpha\beta_n - \gamma_n)\omega_k^5 +(\hat{\beta}_m\gamma_n - \alpha\delta_n -\hat{\gamma}_m\beta_n)\omega_k^3 + (\hat{\gamma}_m\delta_n - \hat{\delta}_m\gamma_n)}{\beta_n\omega_k^6 + (\alpha\gamma_n - \hat{\beta}_m\beta_n-\delta_n)\omega_k^4 + (\hat{\beta}_m\delta_n + \hat{\delta}_m\beta_n -\hat{\gamma}_m\gamma_n)\omega_k^2 -\hat{\delta}_m\delta_n}\right) \Bigg].}
\end{array}
\end{equation}
This allows us to define the following:
\begin{equation}\label{tauzero}
\displaystyle{\tau_n^* = \tau_{n,k_0}^{(\xi_0)}=\min_{1\le k\le 4,\;\xi\ge 1}\{\tau_{n,k}^{(\xi)}\},\quad \omega_0 = \omega_{k_0}.}
\end{equation}

In order to establish whether the steady state $E_j$, $j = 1,2,3$, actually undergoes a Hopf bifurcation at $\tau_n=\tau_n^*$, one has to compute the sign of $d[\operatorname{Re}\lambda(\tau_n^*)]/d \tau_n$. Differentiating the equation (\ref{eq:chp_single_delay}) with respect to $\tau_n$ yields
\[
\left(\frac{d\lambda}{d\tau_n}\right)^{-1} = \frac{(4\lambda^3 + 3\alpha\lambda^2 + 2\hat{\beta}_m\lambda + \hat{\gamma}_m)e^{\lambda\tau_n}  +   2\beta_n\lambda + \gamma_n}{\lambda (\beta_n\lambda^2 + \gamma_n\lambda + \delta_n)} - \frac{\tau_n}{\lambda}.
\]
Introducing the notation $\displaystyle{U = \omega_0^2[\omega_0^2\gamma_n^2 + (\delta_n - \beta_n \omega_0^2)^2]}$, it is clear that $U>0$ for all $\omega_0>0$, and
\begin{equation}
\left(\frac{d\operatorname{Re}\lambda(\tau_n^*)}{d \tau_n}\right)^{-1}\!\!=\frac{1}{U}\!\! \left[\underbrace{A\cos(w_0\tau_n)+B\sin(w_0\tau_n)}_{:=\Gamma} + 2\beta_nw_0^2(\delta_i - \beta_nw_0^2) -\gamma_n^2w_0^2\right]\!\!, 
\end{equation}
where
\[
\begin{array}{l}
A =2\omega_0^2(\hat{\beta}_m - 2\omega_0^2)F_2 + \omega_0(3\alpha \omega_0^2 - \hat{\gamma}_m)F_1,\\\\ 
B=-\omega_0(3\alpha \omega_0^2 - \hat{\gamma}_m)F_2  + 2\omega_0^2( \hat{\beta}_m-2\omega_0^2 )F_1,\\\\
\Gamma= 2\omega_0^2(\hat{\beta}_m - 2\omega_0^2)H_1 + \omega_0(3\alpha\omega_0^2 - \hat{\gamma}_m)H_2.
\end{array}
\]
Consequently, with $v_0 = {w_0}^2$ one can write $d[\operatorname{Re}\lambda(\tau_n^*)]/d \tau_n$ as follows
\begin{equation}
\begin{array}{l}
 \displaystyle{\left(\frac{d\operatorname{Re}\lambda(\tau_n^*)}{d \tau_n}\right)^{-1}=
\frac{1}{U}\left[4{w_0}^8 + 3c_3{w_0}^6+2c_2{w_0}^4+c_1{w_0}^2\right]}\\\\
 \displaystyle{=\frac{1}{U}\left[4{v_0}^4 + 3c_3{v_0}^3+2c_2{v_0}^2+c_1{v_0}\right]=\frac{v_0}{U}h'(v_0),}
\end{array}
\end{equation}
where $h(v)$ is defined in (\ref{equation:h(v)}). Since $v_0 = {w_0}^2>0$, this implies
\[
\begin{array}{l}
\displaystyle{\mathrm{sign}\left(\frac{d\operatorname{Re}\lambda({\tau_n}^*)}{d \tau_n}\right) =  \mathrm{sign}\left[\left(\frac{d\operatorname{Re}\lambda({\tau_n}^*)}{d \tau_n}\right)^{-1}\right]}\\\\
\displaystyle{\hspace{1cm}=\mathrm{sign}\left[\frac{v_0h'(v_0)}{U}\right]=\mathrm{sign} \left[h'(v_0)\right].}
\end{array}
\]
We can therefore conclude the following result.

\begin{theorem}\label{Theorem:Stability}Let the coefficients of the characteristic equation at the steady state $E_j$, $j=1,2,3$, with $\tau_{1,2}=0$, be given by (\ref{sys:nodelays}). Suppose these coefficients satisfy the Routh-Hurwitz criteria, namely, $p_0,\ldots,p_4>0$, $p_3p_2>p_1p_4$, and $p_3p_2p_1>p_4p_1^2 + p_3^2p_0$. Additionally, let $\omega_0$ and $\tau_n^*$, $n=1,2$, be defined as in (\ref{tauzero}) with $h'(\omega_0^2)> 0$, where $\tau_m = 0$ for $m\ne n$ . Then, the steady state $E_j$ of the system (\ref{system:garbage}) is stable for $\tau_n<\tau_n^*$, unstable for $\tau_n>\tau_n^*$, and undergoes a Hopf bifurcation at $\tau_n=\tau_n^*$.
\end{theorem}

\begin{remark}
The {\bf Theorem \ref{Theorem:Stability}} only holds if the quartic (\ref{equation:h(v)}) has at least one positive real root, which is guaranteed in the special case of $c_0<0$. However, when $c_0\ge 0$, it is impractical to consider the analytical distribution of roots. Hence, one would have to compute these roots numerically to verify the assumptions of the theorem.
\end{remark}

\subsection{Garbage- and mRNA-associated amplification delays are non-zero}

Let us now consider the most complex situation where both time delays $\tau_1$ and $\tau_2$ associated with the primed amplification are positive. In this case the characteristic equation (\ref{sys:nodelays}) can be rearranged into the following equation
\begin{equation}\label{eq:chp_two_delays}
\sigma(\lambda) =  \sigma_0(\lambda) + \sigma_1(\lambda)e^{-\lambda \tau_1} + \sigma_2(\lambda)e^{-\lambda \tau_2} = 0,
\end{equation}
where
\[
\begin{array}{l}
\sigma_0 = [\sigma_{01} (\lambda+ d_g) +\sigma_{02}](\lambda+ a) +  a b p n_2 S M^2(\lambda + d_g),\\\\
\sigma_1 =  a b_2 n_2M S(b_3S +\lambda+d_g) (bM S - \lambda M-\hat{h}),\\\\
\sigma_2 = -ab_{{3}}n_{{2}}S\left[ -bb_{{1}}n_{{3}}{M}^{2}{S}^{2} + \left( b_{{3}}S+\lambda+2\,d_{{g}} \right)  ( \lambda M+\hat{h})G\right],\\\\
\sigma_{01} = -{S}^{2}{M}^{2}+ \left(  D a\lambda\,n_{{2}}+S{\lambda}^{2} \right) M+\hat{h}\lambda S+a\hat{h}n_{{2}}D,\\\\
\sigma_{02} = -bb_3 ( b_1 n_3 +b) M^2 + b_3(an_2D + d_gG)(\lambda M + \hat{h})S + b_3\lambda(\lambda M + \hat{h}S )S^2.
\end{array}
\]
To analyse the distribution of roots of the equation (\ref{eq:chp_two_delays}) we follow the methodology introduced by Gu {\it et al.} \cite{Gu2005} and subsequently used for analysis of other systems with multiple time delays \cite{NKB16a,BKHS08}. Let $T$ denote the {\it stability crossing curves} which is the set of all the {\it crossing points }  $(\tau_1,\tau_2)\in\mathbb{R}^2_+ $, for which the characteristic polynomial  $\sigma(\lambda)$ has at least one purely imaginary root.  Introducing the parameterisation
\begin{equation}
\delta_k(\lambda) = \frac{\sigma_j(\lambda)}{\sigma_0(\lambda)}, \qquad k = 1,2,
\end{equation}
the equation (\ref{eq:chp_two_delays}) transforms into
\begin{equation}
\delta(\lambda,\tau_1,\tau_2) = 1 + \delta_1 (\lambda) e^{-\lambda\tau_1} + \delta_2(\lambda)e^{-\lambda\tau_2}=0.
\end{equation}
It is important to note that this parameterisation is only valid as long as $\sigma_0$ does not have any imaginary zeros. Hence, by Proposition 3.1 in \cite{Gu2005}, for each $\omega\ne 0$, $\lambda = i\omega$ is a solution of $\sigma(\lambda,\tau_1,\tau_2) = 0$ for some $(\tau_1,\tau_2)\in \mathbb{R}^2_+$ if and only if
\begin{enumerate}[(i)]
	\item Given  $\sigma_0(i\omega)\ne0$
\begin{gather}\label{cond:2Delays_1}
	\begin{aligned}
	&|\delta_1(i\omega)| + |\delta_2(i\omega)|\ge 1,\\
	-1\le &|\delta_1(i\omega)| - |\delta_2(i\omega)|\le 1.
	\end{aligned}
	\end{gather}
	\item Given  $\sigma_0(i\omega)  = 0$,
	\begin{equation}
	|\sigma_1(i\omega)| = |\sigma_2(i\omega)|.
	\end{equation}
\end{enumerate}
Let $\Omega$ denote the {\it crossing set}, i.e the set of all $\omega>0$ which satisfy the conditions (i),(ii) above. This set consists of $N$ intervals with a finite length. Moreover, if the intervals are ordered such that the left-end point of $\Omega_k$ is an increasing function of $k$, $k = 1,2,...,N$, then we have that 
\begin{equation}
\Omega = \bigcup_{k = 1}^N \Omega_k.
\end{equation}
Thus, for any given $\omega\in\Omega$ satisfying $\sigma_j(i\omega)\ne 0$, $j = 0,1,2$, the critical time delay pairs satisfying $\sigma(\lambda,\tau_1,\tau_2) = 0$ with $\lambda=i\omega$ are given by
\begin{equation}
\begin{array}{l}
(\tau_1^*,\tau_2^*)\in \mathcal{T}= \left\{\mathcal{T}_\omega|\omega\in \Omega\right\},\\\\
\mathcal{T}_\omega = \left( \bigcup_{u\ge u^+_0, v\ge v^+_0} \mathcal{T}^+_{\omega,u,v} \right) \cup\left( \bigcup_{u\ge u^-_0, v\ge v^-_0}
\mathcal{T}^-_{\omega,u,v}\right),
\end{array}
\end{equation}
where
\begin{equation}
\mathcal{T}^{\pm}_{\omega,u,v} = \left\{ (\tau^{u\pm}_1,\tau^{v\pm}_2)\right\},
\end{equation}
with
\begin{equation}\label{stab_curves}
\begin{array}{l}
\displaystyle{\tau_1 = \tau^{u\pm}_1(\omega) = \frac{\mathrm{Arg}[\delta_1(i\omega)] + (2u - 1)\pi \pm \theta_1}{\omega}\ge 0 , \quad u = u^{\pm}_0, u^{\pm}_0 + 1, u^{\pm}_0 +2,...}\\\\
\displaystyle{\tau_2 = \tau^{v\pm}_2(\omega) = \frac{\mathrm{Arg}[\delta_2(i\omega)] + (2v - 1)\pi \mp \theta_2}{\omega}\ge 0 , \quad v = v^{\pm}_0, v^{\pm}_0 + 1, v^{\pm}_0 +2,...}
\end{array}
\end{equation}
and the angles $\theta_{1,2}\in[0,\pi]$ are computed as follows
\begin{equation}
\begin{array}{l}
\displaystyle{\theta_1 = \arccos\left( \frac{1 + |\delta_1(i\omega)|^2 - |\delta_2(i\omega)|^2}{2|\delta_1(i\omega)|}\right),}\\\\
\displaystyle{\theta_2 = \arccos\left( \frac{1 + |\delta_2(i\omega)|^2 - |\delta_1(i\omega)|^2}{2|\delta_2(i\omega)|}\right),}
\end{array}
\end{equation}
where $u^{\pm}_0$ and $v^{\pm}_0$ are the smallest possible integers for which the corresponding delays $\tau^{u^{\pm}_0\pm}_1,\; \tau^{v^{\pm}_0\pm}_2$ are non-negative.

\section{Numerical stability analysis and simulations}

\begin{table}
	\scriptsize
	\begin{tabular}{|  p{.15\linewidth}|   p{.45\linewidth}| p{.1\linewidth}| p{.17\linewidth}|}
		\hline
		Parameter & Biological meaning & Value & Units \\
		\hline
		$d_m$ & mRNA decay rate & $0.14$ & $hr^{-1} $ (half life of 5h)\\
		\hline
		$d_s$ & siRNA decay rate & $2$ & $hr^{-1}$ (half life of 21 mins)\\
		\hline
		$d_g$ & Garbage RNA decay rate & $2.8$ & $hr^{-1}$ (half life of 15 mins)\\
		\hline
		$h$ & mRNA transcription rate & $160$ & $hr^{-1}\;\;cell^{-1}$\\
		\hline
		$p$ & Rate of dsRNA synthesis from RNA & $0.002$ & $hr^{-1}$\\
		\hline
		$a$ & Rate of dsRNA cleavage by Dicer & $2$ & $hr^{-1}$\\
		\hline
		$b_1$ & Rate of RISC-mRNA complex formation & $8\times 10^{-4}$ & $ cell\;mol^{-1}\;hr^{-1}$\\
		\hline
		$b_2$ & Rate of RdRp-mRNA complex formation & $8\times 10^{-5}$ & $ cell\;mol^{-1}\;hr^{-1}$\\
		\hline
		$b_3$ & Rate of RdRp-garbage complex formation & $9\times 10^{-4}$ & $ cell\;mol^{-1}\;hr^{-1}$\\
		\hline
		$n_1$ & Transgene copy number  & $1$ & \\
		\hline
		$n_2$ & Yield of siRNA per cleaved dsRNA  & $10$ & \\
		\hline
		$n_3$ & Yield of garbage RNA from degraded mRNA  & $1$ & \\
		\hline
		$\tau_1$ & Delay in dsRNA synthesis from mRNA & $0$ & \\
		\hline
		$\tau_2$ & Delay in dsRNA synthesis from aberrant RNA & $0$ & \\
		\hline
		
	\end{tabular}
	\caption{Baseline parameter values for the system (\ref{system:garbage}). The majority of the parameter values are taken from \cite{Groenenboom2005}.}
	\label{tab:param}
\end{table}

In order to understand the effects of different parameters on feasibility and stability of different steady states and investigate the role of the time delays associated with primed amplification, we have used a pseudospectral method implemented in a traceDDE suite for MATLAB \cite{breda} to numerically compute the eigenvalues of the characteristic equation (\ref{eq:chp_two_delays}). The baseline parameter values are mostly taken from \cite{Groenenboom2005} and are shown in the Table~\ref{tab:param}. It is assumed that mRNA is stable with a half-life of 5 hours, garbage RNA decays 20 times faster than mRNA, and the half-life of siRNA is taken to be 21 mins as measured in human cells \cite{Chiu2003}. The rest of the baseline parameters are chosen such as to illustrate all the different types of dynamical behavior that the model (\ref{system:garbage}) can exhibit. Since RNA interference is a very complex multi-component process, many parameter values are case-specific and hard to obtain experimentally \cite{Melnyk2011,Liang2012, Himber2015}. Hence, rather than focus on a specific set of parameters, we explore the dynamics through an extensive bifurcation analysis.

\begin{figure}[H]
		\centering
	\includegraphics[width=0.9\linewidth]{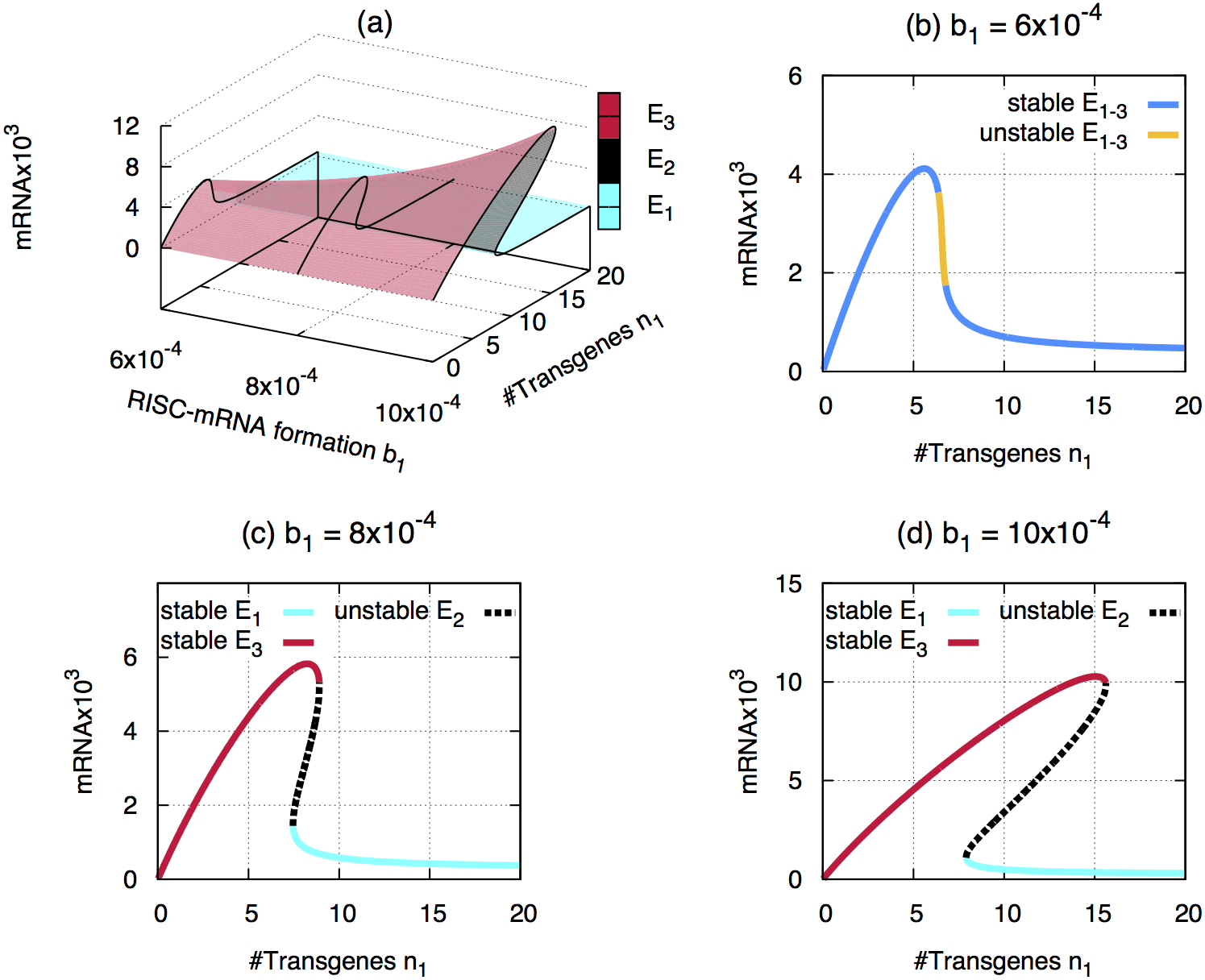}
	\caption{Stability of the steady states $E_1$, $E_2$ and $E_3$ depending on the rate $b_1$ and the number of transgenes $n_1$, with other parameter values taken from Table~\ref{tab:param}.}
	\label{fig:2}
\end{figure}

Figure~\ref{fig:2}(b) shows that if the rate $b_1$, at which the RISC-mRNA complex is formed, is sufficiently small, then only a single steady state $E_{1-3}$ is feasible, and it is stable for small or high numbers of transgenes, and unstable for intermediate values of $n_1$. As the value of $b_1$ increases, sustained silencing occurs at higher numbers of transgenes and higher mRNA levels. The system also acquires an additional unstable feasible steady state $E_2$ with an intermediate level of mRNA, thus creating a region of bi-stability, as shown in Figs.~\ref{fig:2}(c) and (d). The range of values of transgenes $n_1$, for which the bi-stability is observed, itself increases with $b_1$, which means that if the RISC complexes are more efficient in cleaving mRNA (RISC overexpression), it is possible to have the stable states with high and low values of mRNA for higher and lower numbers of transgenes, respectively, and that the range of transgenes for which introduction of dsRNA triggers sustained silencing becomes larger.

A very interesting and counter-intuitive observation from Figs.~\ref{fig:2}(c) and (d) is that the actual values of the steady state mRNA concentration are also growing with $b_1$. One possible explanation for this is that the reduced availability of mRNA means that a smaller amount of it can be directly used to synthesize dsRNA, as described by the $pM(t)$ term in the second equation of (\ref{system:garbage}), and more mRNA is directly degraded into the garbage RNA, thus generating a smaller feedback loop in the model for sufficient silencing to occur.

\begin{figure}[H]
		\centering
	\includegraphics[width=0.9\linewidth]{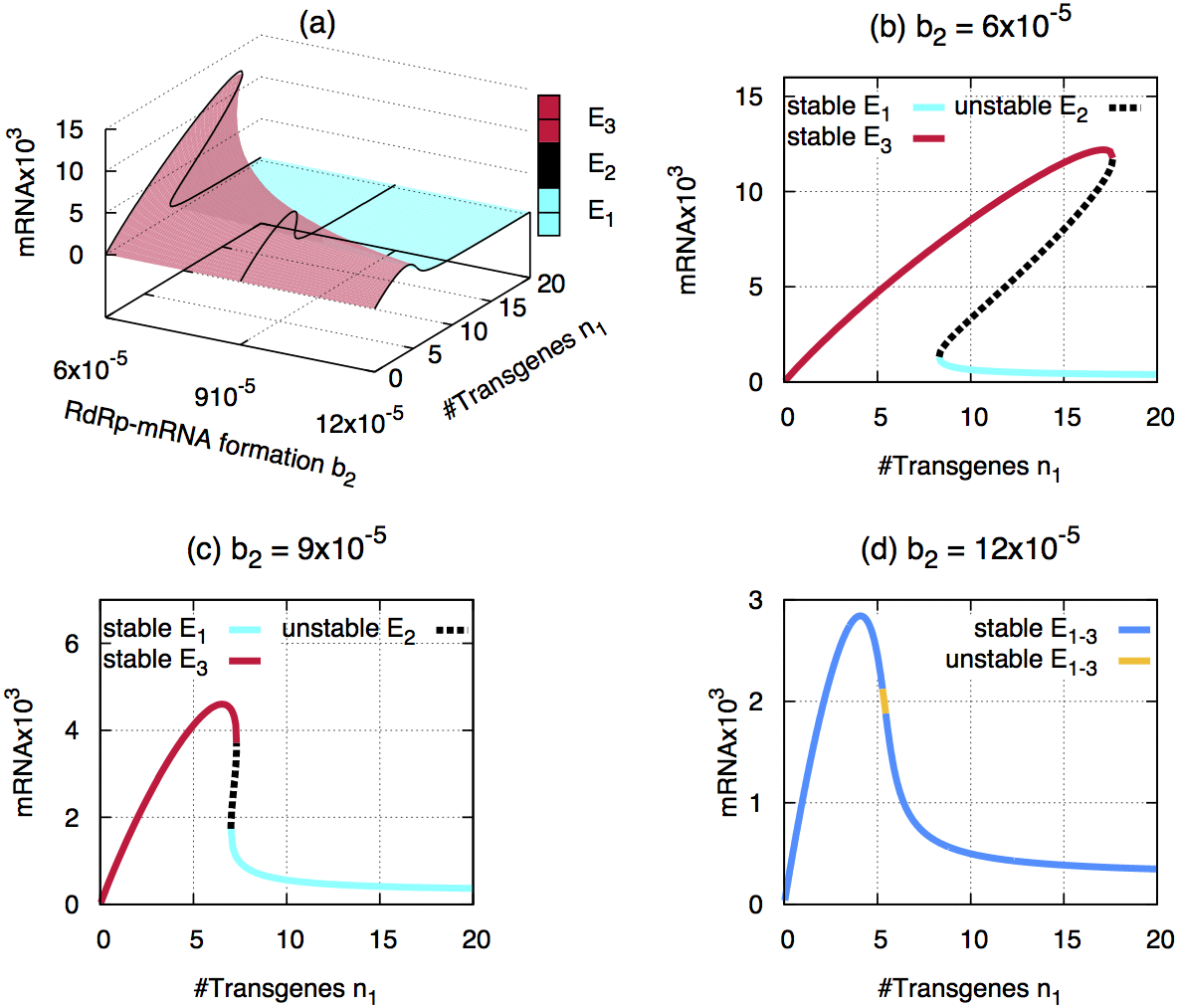}
	\caption{Stability of the steady states $E_1$, $E_2$ and $E_3$ depending on the rate $b_2$ and the number of transgenes $n_1$, with other parameter values from Table~\ref{tab:param}.}
	\label{fig:3}
\end{figure}

\begin{figure}[H]
		\centering
	\includegraphics[width=0.9\linewidth]{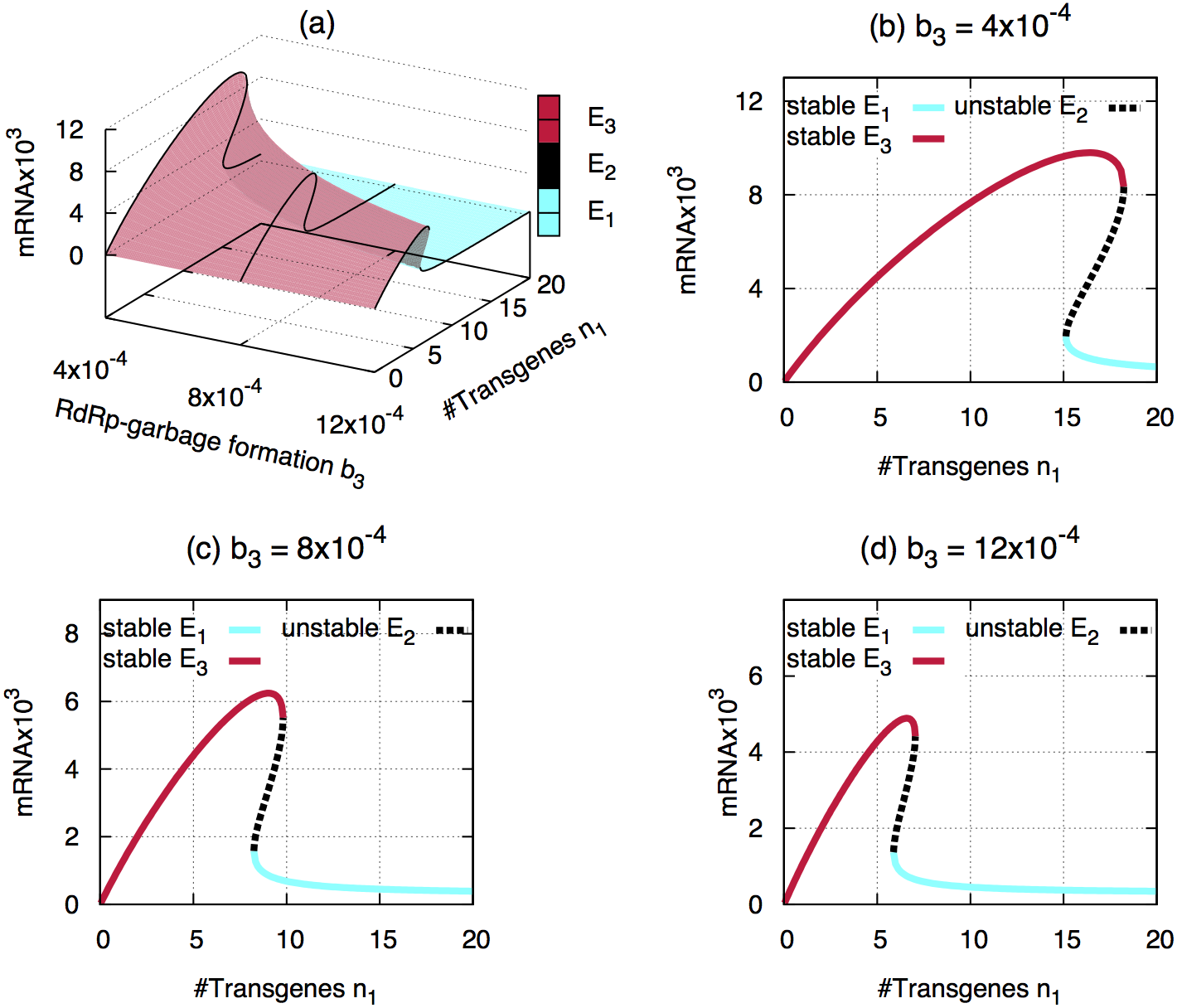}
	\caption{Stability of the steady states $E_1$, $E_2$ and $E_3$ depending on the rate $b_3$ and the number of transgenes $n_1$, with other parameter values from Table~\ref{tab:param}.}
	\label{fig:4}
\end{figure}

When one considers the effect of varying the rate $b_2$ of forming RdRp-mRNA complexes, the behaviour is qualitatively different in that increasing $b_2$ leads to the reduction in the size of the bi-stability region, and for sufficiently high values of $b_2$, the intermediate steady state $E_2$ completely disappears, and the system possesses a single feasible steady state $E_{1-3}$, which is stable for low and high numbers of transgenes, and unstable for intermediate values of $n_1$, as shown in Fig.~\ref{fig:3}. Increasing the rate $b_2$ leads to a decrease in the maximum values that can be attained by the mRNA concentration. Similar behaviour is observed in Fig.~\ref{fig:4}, where the rate $b_3$ of forming RdRP-garbage complexes is varied. Increasing this rate $b_3$ results in a reduced region of bi-stability and smaller values of the maximum mRNA concentration, but at the same time, it does not result in the complete disappearance of the bi-stability region, as was the case when the rate $b_2$ was varied.

\begin{figure}[H]
		\centering
	\includegraphics[width=\linewidth]{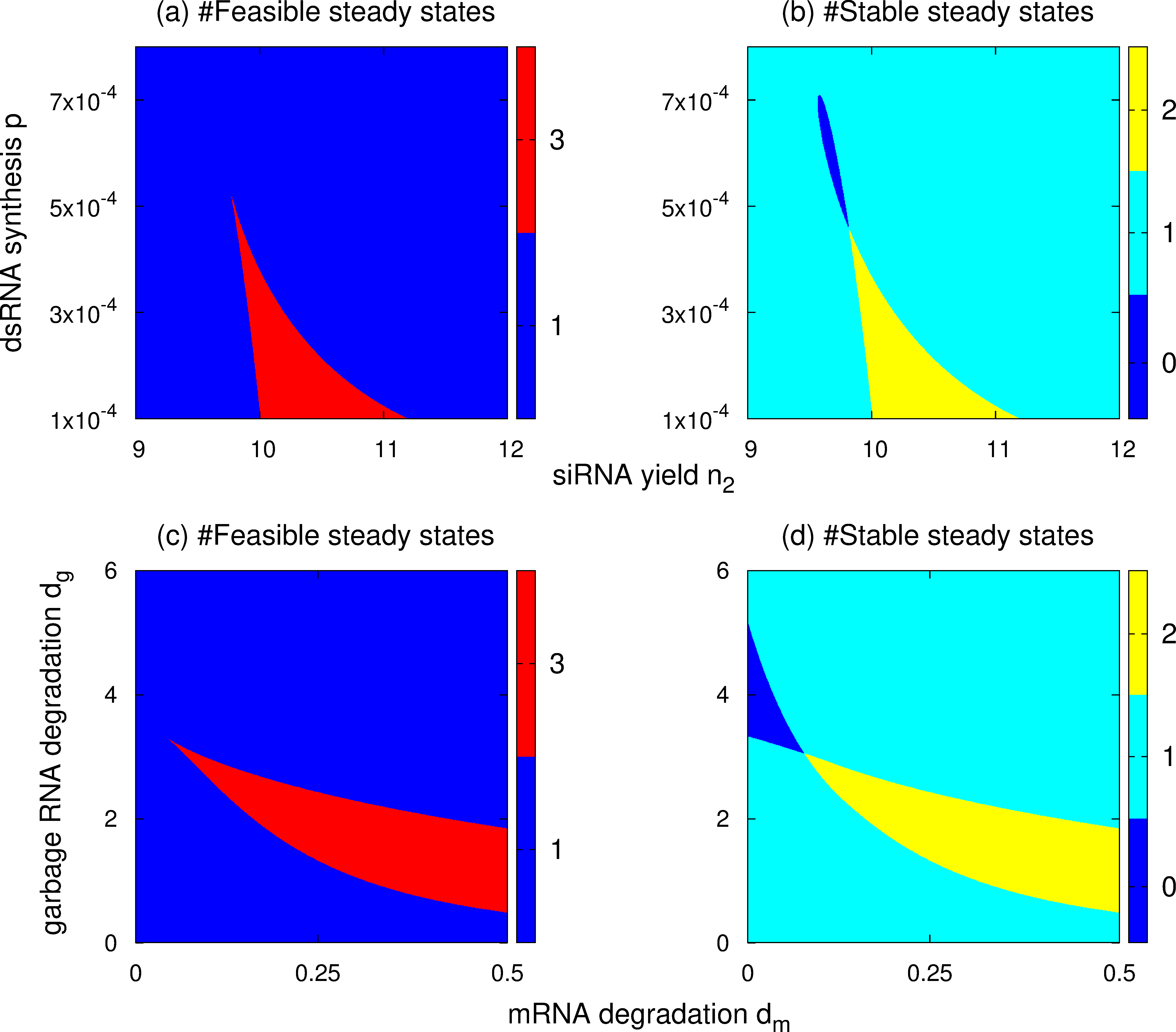}
	\caption{Regions of feasibility and stability of different steady states depending on the degradation rate of mRNA $d_m$, and varying a second parameter as shown by the vertical axis. Other parameter values are taken from Table~\ref{tab:param}.}
	\label{fig:59}
\end{figure}

Comparing the influence of the rate $p$, at which RdRp  synthesises dsRNA directly from the mRNA, to the number of siRNA $n_2$ produced by Dicer per cleaved dsRNA, one can notice that for sufficiently small $n_2$ and $p$, only the steady state $E_3$ is feasible and stable, and, therefore, the strength of RNA silencing is severely limited, with a relatively high concentration of mRNA surviving, as illustrated in Figs.~\ref{fig:59}(a)-(b). This agrees very well with experimental observations in which plants carrying a mutation in RdRp cannot synthesize trigger-dsRNA directly from mRNA, and, thus, fail to induce transgene-induced silencing \cite{Dalmay00}, but similarly to mammals who do not carry RdRp, might experience transient silencing \cite{Caplen01}. Increasing $p$,  reduces the range of $n_2$ values, for which bi-stability occurs, and eventually it leads to the complete disappearance of the intermediate steady state $E_2$. For higher value of $p$, the state $E_{1-3}$ can exhibit instability in a small range of $n_2$ values, and for even higher rates of dsRNA production, this steady state is always stable, thus signifying that gene silencing has been achieved. From a biological perspective, this should be expected, as by increasing $p$, more mRNA can be used for dsRNA synthesis, which is then used for the production of siRNA, which in turn amplifies the process even further. This is consistent with experimental observations which show that  strains of the fungus {\it Neurospora crassa}, which overexpress RdRp, are able to progressively carry fewer transgenes without reverting back to their wild type. As such, even a single transgene is sufficient to induce gene silencing and thus preserve the phenotypic stability of the species \cite{Forrest2004}.

 When one considers the relative effects of the degradation rates of mRNA $d_m$ and garbage RNA $d_g$, it becomes clear that if the mRNA decays quite slowly, while garbage RNA decays fast, in a certain range of $d_g$ values the system does not converge to any steady states but rather exhibits periodic solutions, as shown Figs.~\ref{fig:59}(a)-(b). As the rate of mRNA degradation is increased, this reduces the range of possible $d_g$ values where periodic behaviour is observed, until it eventually disappears completely. It is important to note that higher values of $d_g$ correspond to $E_3$ and lower values correspond to $E_1$, which suggests that decreasing the rate $d_g$ of garbage RNA degradation results in more of it being available for additional dsRNA synthesis, which subsequently results in a more efficient gene silencing.

\begin{figure}[H]
		\centering
	\includegraphics[width=\linewidth]{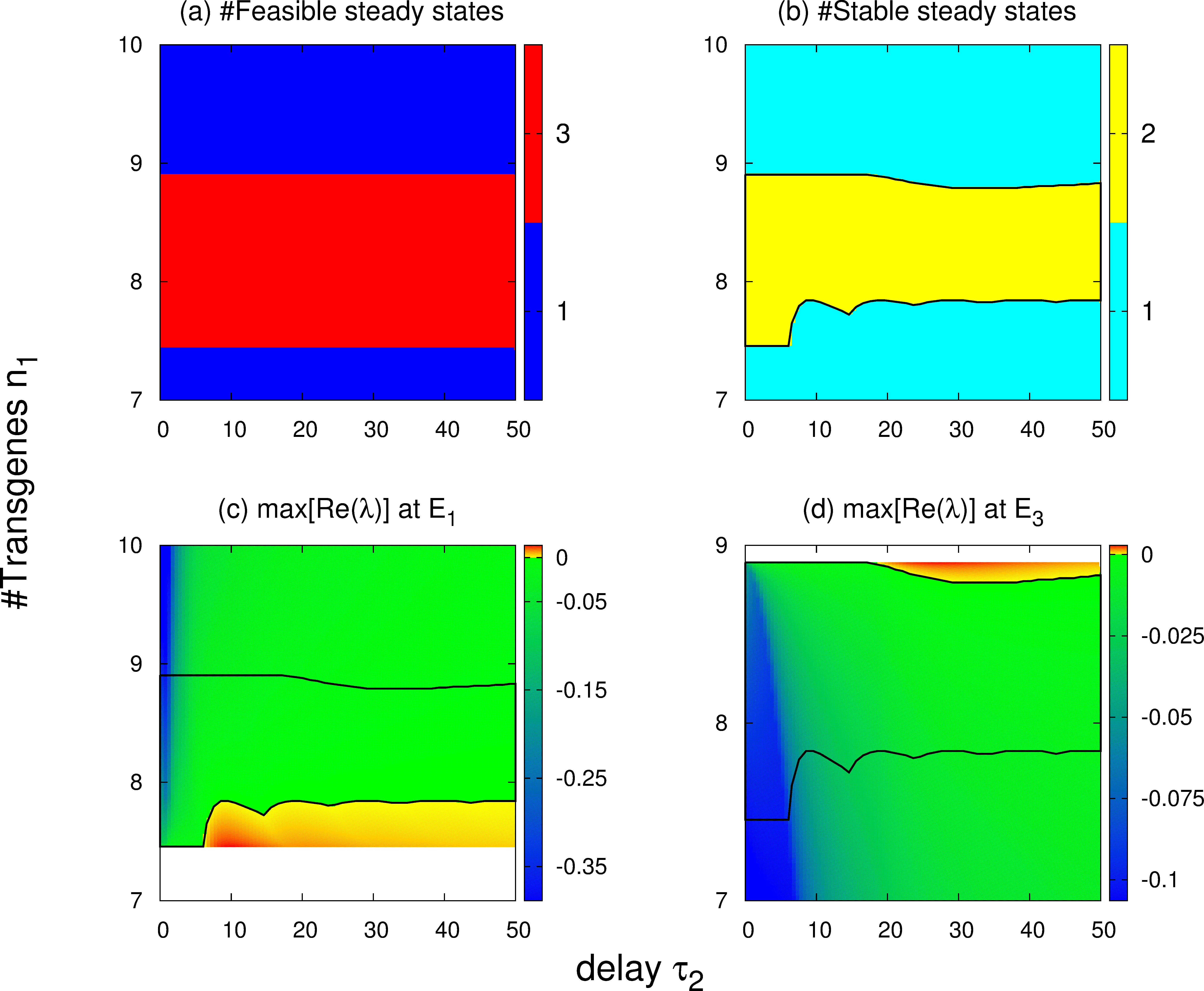}
	\caption{The top row shows the number of feasible (a) and stable (b) steady states depending on the time delay $\tau_2$ and the number of transgenes $n_1$, with $\tau_1=1$, and the rest of the parameter values taken from Table~\ref{tab:param}. The bottom row shows $\max[{\rm Re}(\lambda)]$ for the steady states $E_1$ (c) and $E_3$ (d) with a low and high concentration of mRNA, respectively, while the steady state $E_2$, which has a medium mRNA concentration, is unstable everywhere.}
	\label{fig:10}
\end{figure}

\begin{figure}[H]
		\centering
	\includegraphics[width=\linewidth]{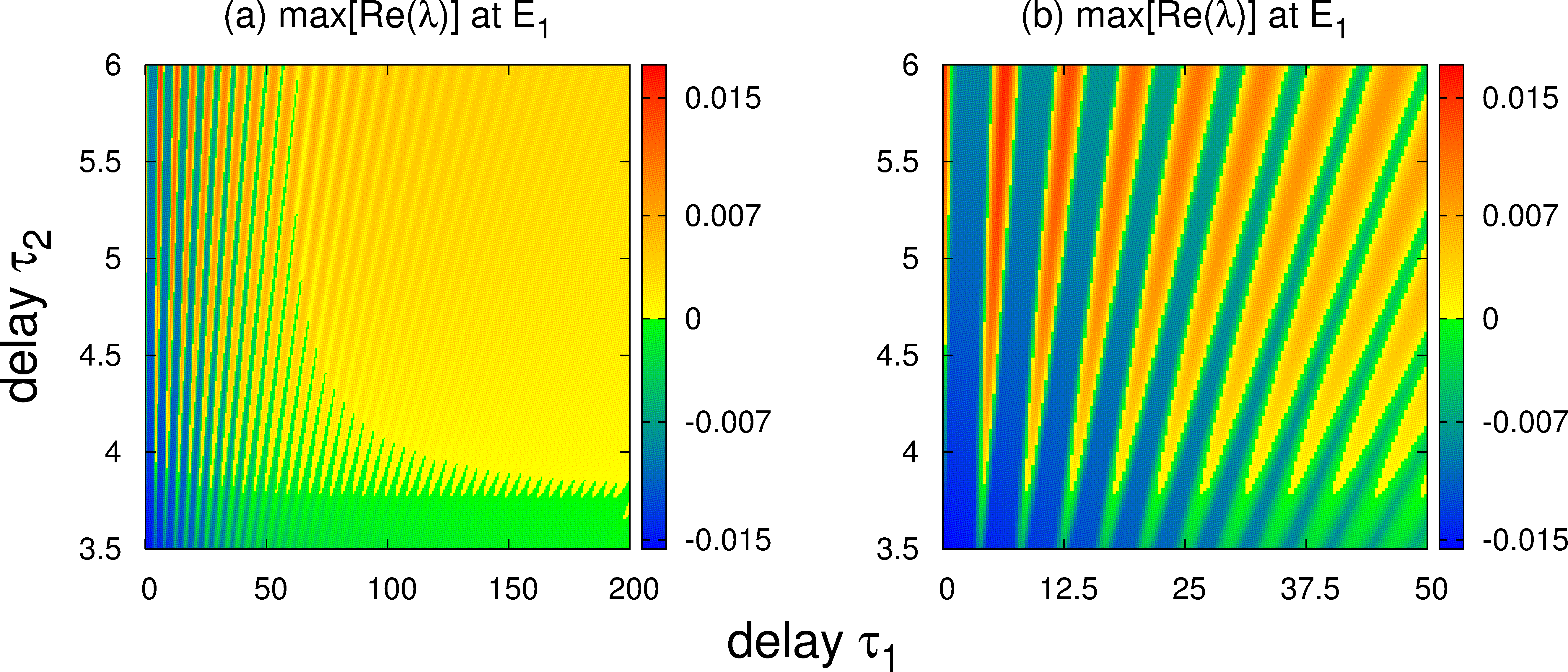}
	\caption{Colour code denotes $\max[{\rm Re}(\lambda)]$ for the steady state $E_1$ with a low concentration of mRNA depending on the two time delays $\tau_1$ and $\tau_2$ associated with primed amplification, with the rest of the parameter values taken from Table~\ref{tab:param}. In the regions where $E_1$ is stable, the system is actually bi-stable, as the steady state $E_3$ with a high  mRNA concentration is also stable.}
	\label{fig:11}
\end{figure}

\begin{figure}[H]
		\centering
	\includegraphics[width=\linewidth]{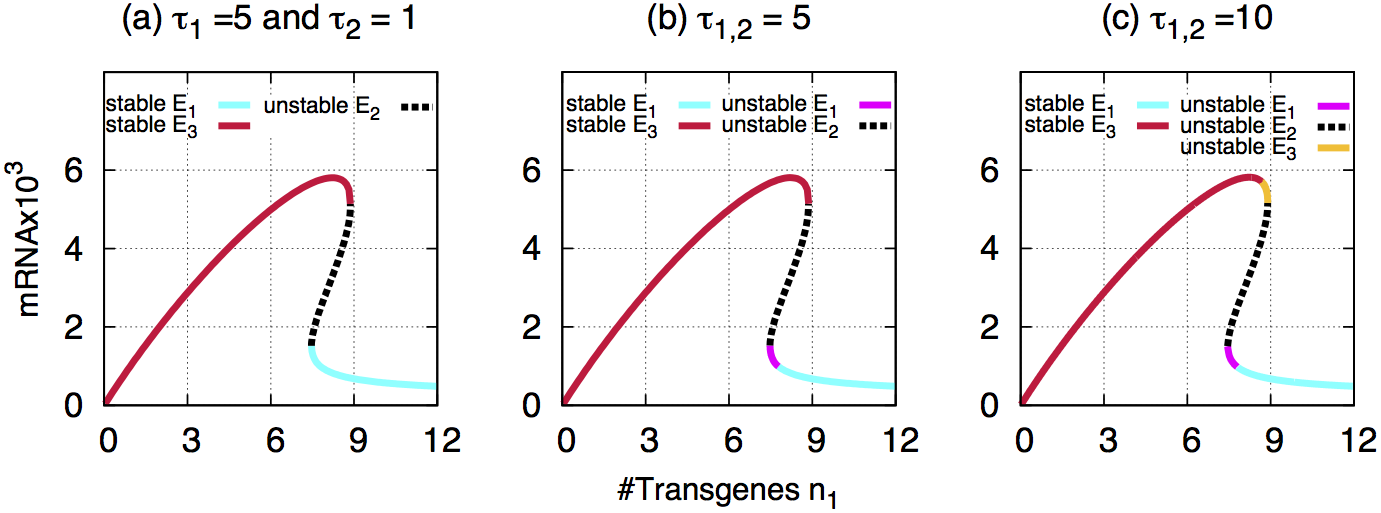}
	\caption{Stability of the three  steady states $E_1$, $E_2$ and $E_3$ with parameter values from Table~\ref{tab:param}. The red and cyan lines denote the regions where the steady states with a low ($E_1$) and high ($E_3$) levels of mRNA are stable, respectively. The black line signifies the steady state $E_2$ with a medium concentration of mRNA which is always unstable. The violet and light-brown lines denote the regions where the steady states $E_1$ and $E_3$ are unstable, respectively.}
	\label{fig:14}
\end{figure}

Figure~\ref{fig:10} shows how the region where the system (\ref{system:garbage}) is bi-stable depends on the number of transgenes and the time delay $\tau_2$, associated with a delayed production of dsRNA from aberrant RNA when the  delay associated with production of dsRNA from mRNA is fixed at $\tau_1 = 1$. This figure shows that when $\tau_2 = 0$, the system is bi-stable in the approximate range $7.5\le n_1\leq 8.9$, and for sufficiently small $\tau_2$ up until $\tau_2\le7$, the behaviour of the system remains largely unchanged, whereas for $\tau_2>7$ and sufficiently small number of transgenes, the silenced steady state $E_1$ loses stability. This stability can be regained for some higher values of $\tau_2$, but then it will be lost again. Steady states $E_1$ with higher values of $n_1$ are not affected by the variations in $\tau_2$ and remain stable throughout the bi-stability region. In a similar way, the steady state $E_3$ can also lose its stability, but unlike $E_1$, this happens for high values of transgenes, and the range of $n_1$ values where instability happens is smaller than for $E_1$. These results suggest that the time delays associated with primed amplification can result in a destabilisation of the steady states $E_1$ and $E_3$, thus disrupting gene silencing. When both time delays are varied, as shown in Figs.~\ref{fig:11} and \ref{fig:14}, the steady state $E_3$ without sufficient silencing is always stable, whereas increasing $\tau_1$ and/or $\tau_2$ causes the silenced steady state $E_1$ to switch between being stable or unstable. We note that the boundaries of the stability crossing curves shown in Fig.~\ref{fig:11} are analytically described by (\ref{stab_curves}). Figure~\ref{fig:14} illustrates that whilst the time delays do not affect the shape of the hysteresis curve, they can cause some extra parts of it to be unstable, which happens for smaller values of the time delay to $E_1$ only, and for higher values of the time delays to $E_3$ as well. A possible interpretation of this result is that the feedback loop in the model is highly sensitive to the speed dsRNA production from its constituent parts. When the dsRNA synthesis is hindered by the time delays, the production cannot maintain the required consistent pace, and, as a result, one of the steady states loses stability, which gives birth to stable periodic solutions.

\begin{figure}[H]
	\centering
	\includegraphics[width=0.85\linewidth]{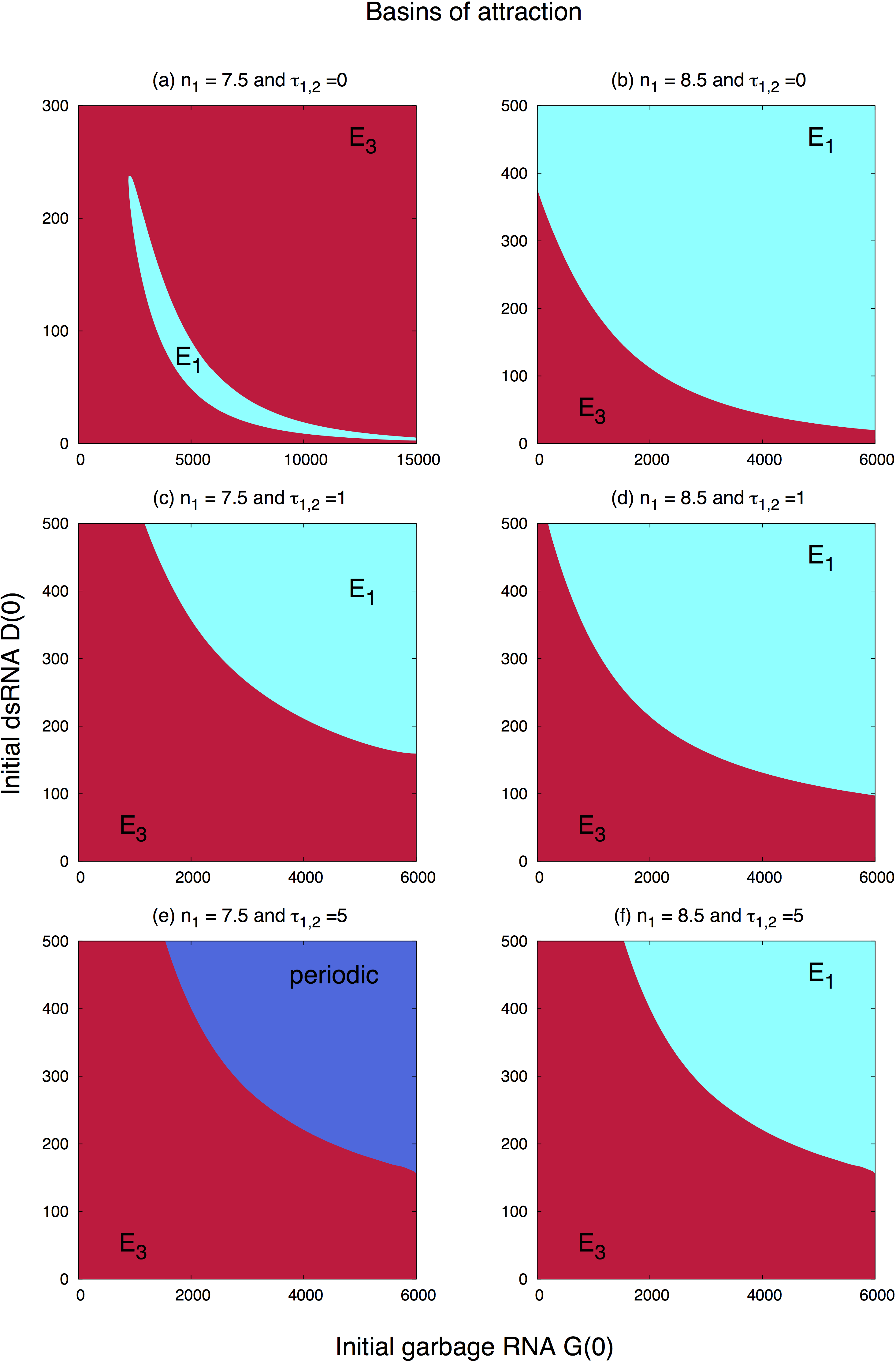}
	\caption{Basins of attraction of different steady stated depending on the initial dosage of dsRNA and garbage RNA within the host cell. The red and cyan regions are where the system converges to the steady state with a high, $E_3$, and low, $E_1$, levels of mRNA, respectively. In the dark-blue region the system exhibits periodic oscillations around the steady state $E_1$.}
	\label{fig:12}
\end{figure}

\begin{figure}[H]
		\centering
	\includegraphics[width=0.85\linewidth]{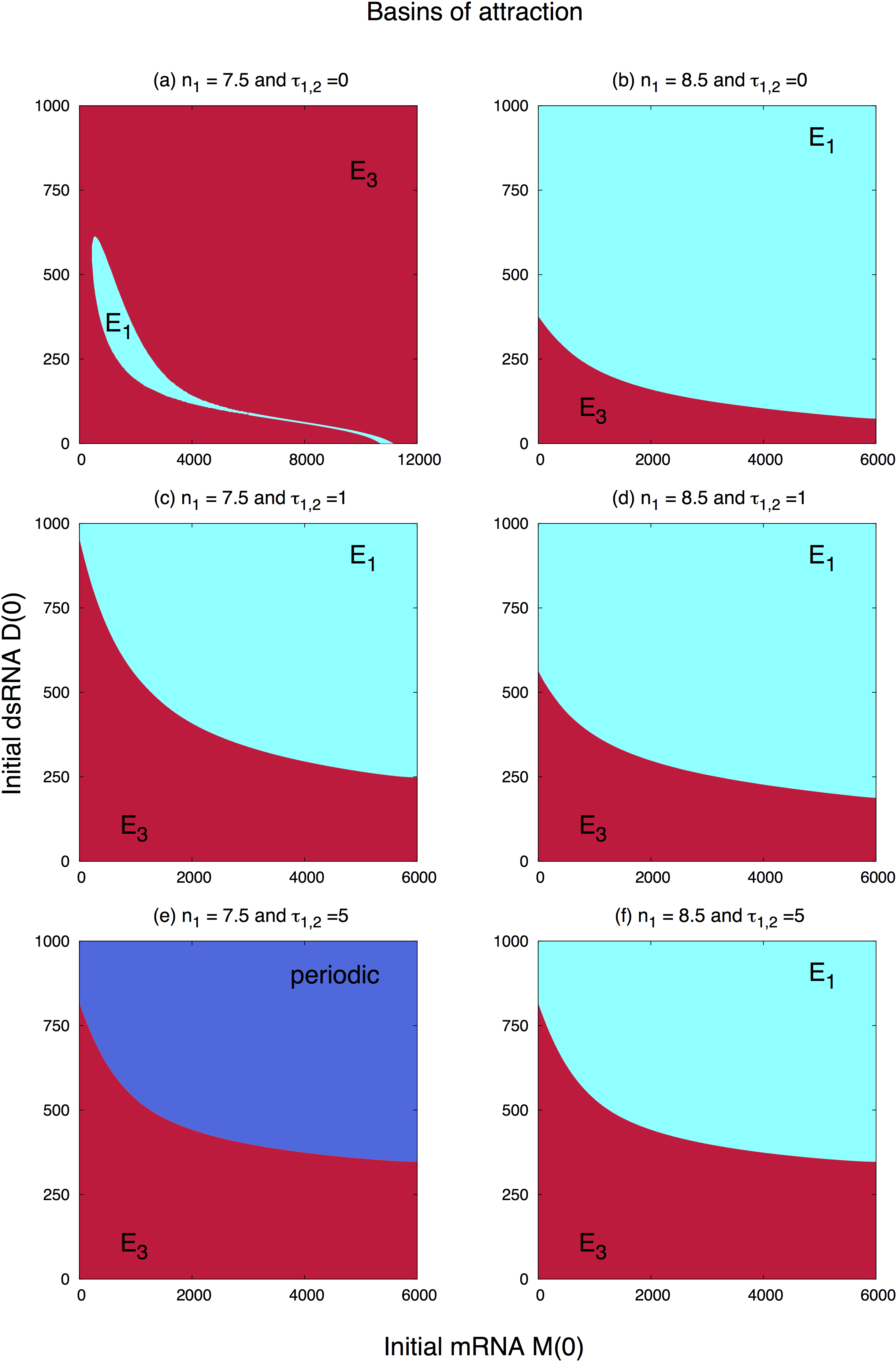}
	\caption{Basins of attraction of different steady stated depending on the initial dosage of dsRNA and initial mRNA within the host cell. The red and cyan regions are where the system converges to the steady state with a high, $E_3$, and low, $E_1$, levels of mRNA, respectively. In the dark-blue region the system exhibits periodic oscillations around the steady state $E_1$.}
	\label{fig:13}
\end{figure}

Figures~\ref{fig:12} and \ref{fig:13} illustrate how the initial dosage of the dsRNA $D(0)$, garbage RNA $G(0)$ and mRNA $M(0)$ affect the behaviour of the model. Starting with the smaller number of transgenes, for which the system (\ref{system:garbage}) is bi-stable we see that in Figs.~\ref{fig:12}(a),(b) and Figs.~\ref{fig:13}(a),(b), when the delays $\tau_1$ and $\tau_2$ are both set to zero, the system mostly converges to the steady state with a relatively high concentration of mRNA $E_3$ for smaller numbers of transgenes $n_1$, and to the steady state with a lower concentration of mRNA $E_1$ for higher numbers of transgenes $n_1$. As the time delays associated with the primed amplification increase, this increases the basin of attraction of $E_1$ for smaller $n_1$, and the basin of attraction of $E_3$ for higher $n_1$, as shown in Figs.~\ref{fig:12},\ref{fig:13}(c) and Figs.~\ref{fig:12},\ref{fig:13}(d), respectively. These figures suggest that for sufficiently high dosage of dsRNA and initial garbage RNA or mRNA being present in the cell, the system achieves a stable steady state where gene silencing is sustained. For higher values of the time delays, there is a qualitative difference in behaviour between lower and higher numbers of transgenes. For lower numbers of transgenes, the system exhibits a bi-stability between a stable steady state $E_3$ with a high concentration of mRNA and a periodic orbit around the now unstable steady state $E_1$. On the other hand, for higher values of $n_1$, there is still a bi-stability between $E_1$ and $E_3$. Whilst in this case, the system may appear not to be as sensitive to the effects of time delays in the primed amplification pathway, it is still evident that in the presence of time delays one generally requires a higher initial dosage of dsRNA to achieve sustained silencing. Furthermore, in the narrow range of $n_1$ values, where the steady state $E_3$ is destabilised by the time delays, numerical simulations show that the system always moves towards a stable steady state $E_1$ rather than oscillate around $E_3$, thus suggesting that the Hopf bifurcation of the steady state is subcritical.

\begin{figure}[H]
		\centering
	\includegraphics[width=\linewidth]{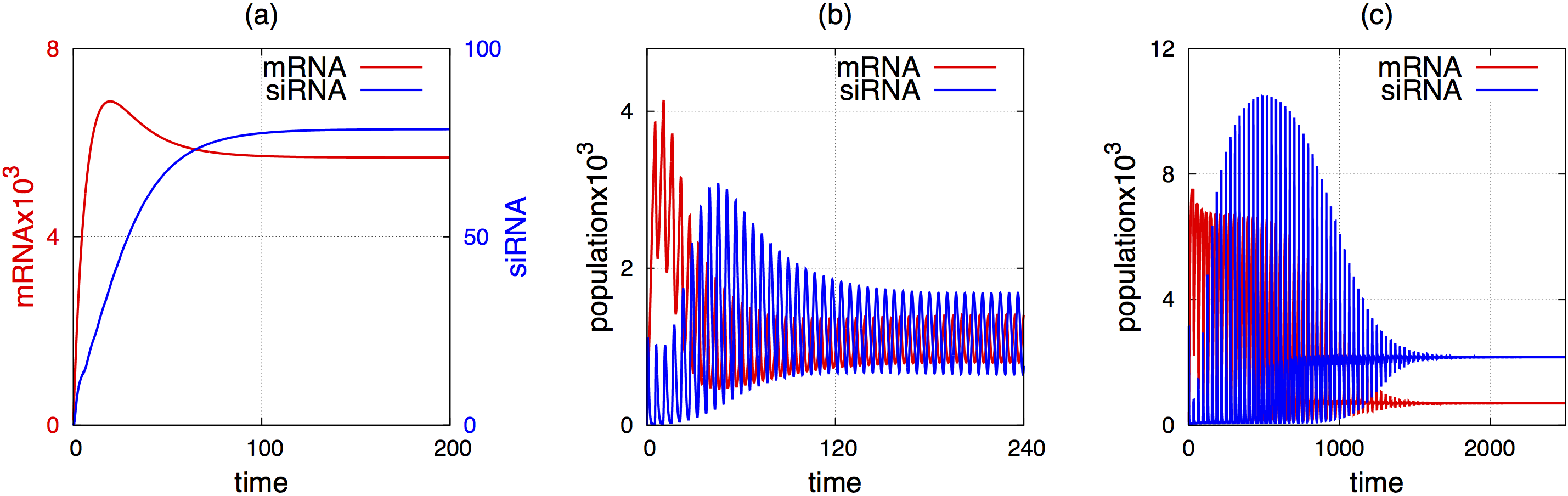}
	\caption{Numerical solutions of the model (\ref{system:garbage}). (a) Stable steady state $E_3$ for $\tau_1=\tau_2=5$, $n_1=7.5$. (b) Periodic oscillations around the steady state $E_1$ for $\tau_1=\tau_2=5$, $n_1=7.5$. (c) Transient oscillations settling on a stable steady state $E_1$ for $\tau_1=1$, $\tau_2=30$ and $n_1=8.87$. Other parameter values are taken from Table~\ref{tab:param}.}
	\label{num_sim}
\end{figure}

To illustrate the dynamics of the system (\ref{system:garbage}) in different dynamical regimes, we have solved this system numerically, and the results are presented in Fig.~\ref{num_sim}. Figures (a) and (b) demonstrate the regime of bi-stability shown in Figs.~\ref{fig:12},\ref{fig:13}(e), where under the presence of both time delays and depending on the initial conditions, the system either approaches the default stable steady state $E_3$ under a low initial dsRNA dosage, or tends to a periodic orbit around the silenced steady state $E_1$ despite a high initial dsRNA dosage. Figure (c) corresponds to a situation where the number of transgenes is sufficiently high, and the steady state $E_3$ is destabilised by the time delays, in which case the system approaches a silenced steady state $E_1$. It is interesting to note that prior to settling on the silenced state $E_1$, the system exhibits a prolonged period of oscillations around this state - a phenomenon very similar to the one observed in models of autoimmune dynamics \cite{BN12,BN15}, where the system can also show oscillations and then settle on some chronic steady state. This behaviour highlights an important issue that during experiments one has to be able to robustly distinguish between genuine sustained oscillations and long-term transient oscillations that eventually settle on a steady state.

\section{Discussion}

In this paper we have considered a model of RNA interference with two primed amplification pathways associated with the production of dsRNA from siRNA and
two separate RdRp-carrying complexes formed by targeting mRNA and garbage RNA. For better biological realism, we have explicitly included distinct time delays for each of these pathways to account for delays inherent in dsRNA synthesis. The system is shown to exhibit up to three biologically feasible steady states, with a relatively low $(E_1)$, medium $(E_2)$, or high ($E_3$) concentration of mRNA. 

Stability analysis of the model has shed light onto relative importance of different system parameters. For sufficiently small levels of host mRNA, the system has a single stable steady state $E_3$, whose mRNA concentration is growing with the number of transgenes $n_1$. Experimental observations suggest that the amount of transcribed mRNA is an important factor in the ability of transcripts to trigger silencing. Production of mRNA can generally be enhanced in two ways: either the target transgene is under control of a 35S promoter with a double enhancer so that the gene is transcribed at a higher rate \cite{Elmayan96}, or there are enough transgenic copies to maintain an adequate production of mRNA to trigger silencing. In our model, the number of trangenes $n_1$ and the transcription rate of mRNA $h$ are qualitatively interchangeable. Hence, as the number of transgenes increases, there is a range of transgenic copies for which the system is bi-stable, exhibiting steady states with a high ($E_3$) and low $(E_1)$ mRNA concentrations, where $E_1$ describes a silenced state. For higher values of $n_1$, only the steady state $E_1$ is feasible and stable, suggesting that a sustained state of gene silencing is achieved. From a biological perspective, it is very interesting and important to note that in the bi-stable region, it is not only the parameters, but also the initial conditions that determine whether RNA silencing occurs. This implies that the dosage of dsRNA, which initialises the RNA interference mechanism, as well as the current levels of mRNA and garbage RNA within the cell, determine the evolution of the system. In the absence of time delays, a high dosage of dsRNA and an initial concentration of mRNA or garbage RNA results in a silenced steady state.

In the case when the delays associated with the primed amplification are non-zero, our analysis shows that for specific range of $\tau_1$ and $\tau_2$, both steady states $E_1$ or $E_3$ can lose stability in the bistable region. Once again, not only the parameters, but also the initial conditions control whether the system will converge to the remaining stable state or will oscillate around the unstable steady state. Additionally, in the presence of time delays, one generally requires an even higher initial dosage of dsRNA to achieve sustained silencing compared to the non-delayed model. Interestingly, oscillations can only happen around the silenced steady state $E_1$, and when the steady state $E_3$ loses its stability, the system just moves towards a stable steady state $E_1$. Oscillations around $E_1$ biologically correspond to switching between higher and lower concentrations of mRNA, implying that at certain moments during time evolution, the exogenous mRNA is silenced, and at other times it is not affected by the RNAi. It follows that this switching behaviour might have case-specific implications for the phenotypic stability of a species, which most likely depends on the amplitude of oscillations around the silenced steady state. The biological significance of this result lies in the fact that there are cases where even a high initial dosage of dsRNA will not always result in a silenced steady state. Thus, the augmented model exhibits an enriched dynamical behavior compared to its predecessor which otherwise can only be replicated by different extensions to the core pathway, like the RNase model developed in \cite{Groenenboom2005}, which assumes the presence of a specific siRNA-degrading RNase with saturating kinetics. An interesting open question is whether the switching behavior could also act as a form of protection against the self-inflicted response to an erroneous distinction of target mRNA, and whether periodic silencing can, to some extent, minimise the damage to the host cell. Another issue is that the time delays considered in the model are assumed to be discrete, and hence it would be very insightful and relevant from a biological perspective to investigate how stability results for this model would change in the case where the time delays obey some distribution. Recent results suggest that distributed delays can in some instances increase \cite{KBS11,KBS13,KBS14}, and in others reduce \cite{Rah15} parameter regions where oscillations are suppressed. Our future research will look into the effects of distributed time delays on primed amplification in RNAi.

\newpage

\begin{figure}[H]
		\centering
	\includegraphics[width=\linewidth]{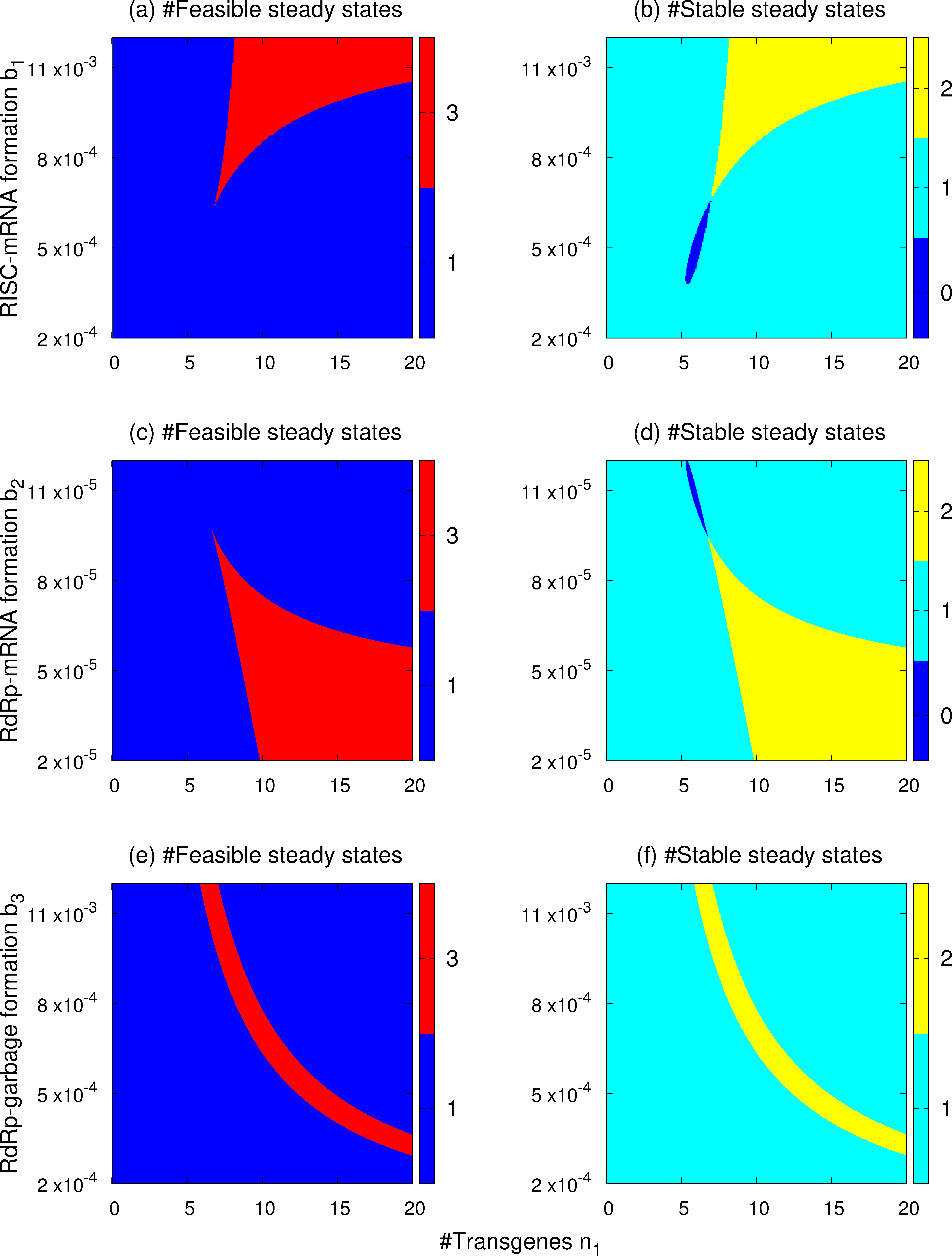}
	\caption{Regions of feasibility and stability of different steady states depending on the number of transgenes $n_1$, and varying one of the complex formation rates as shown by the vertical axis. Other parameter values are taken from Table~\ref{tab:param}.}
	\label{fig:678}
\end{figure}

\newpage

\noindent {\bf Appendix}\\

\renewcommand{\arraystretch}{2}

\begin{table}[h]
	
	\begin{tabular}{| l| l|}
		\hline
		$p_0 = p_{03}S^3 +  p_{02}S^2 + p_{01}S + p_{10}$  & $p_1 = p_{13}S^3 + p_{12}S^2 + p_{11}S + p_{10}$ \\
		\hline
		
		$p_2 = p_{22}S^2 + p_{21}S + p_{20}$ &	$p_4 = MS$	  \\
		\hline
		
		\multicolumn{2}{|l|}{  $p_3 = b_{{3}}M{S}^{2} + \left[ \left( a+d_{{g}} \right) M+n_{{1}}h \right] S+an_{{2}}DM$} \\
		\hline
		
	\end{tabular}
\caption{Coefficients $p_i$, $i=1,...,4$, from the characteristic equation (\ref{sys:nodelays}).}
\label{table:pi}
\end{table}

\begin{table}[h]
	\begin{tabular}{| l| l|}
		\hline
		$p_{00} = n_1n_2ahd_gD$ & $\quad p_{03} = -bb_3M^2[n_3b_1(1-n_2T_2) + b - b_2n_2T_1]$\\
		\hline
		\multicolumn{2}{|l|}{$p_{01} = an_1n_2h[ab_3D - d_g(2b_3T_2G+b_2T_1M)] + ad_g(n_2pbM^2 + n_1hb_3G)$} \\
		\hline
		\multicolumn{2}{|l|}{$p_{02} = bM^2(pn_2b_3 +n_2b_2d_gT_1 - bd_g ) - n_1n_2hb_3(b_2T_1M + b_3T_2G)$}\\
		\hline
	\end{tabular}
\caption{Coefficients $p_{0i}$, $i=0,1,2,3$, from the characteristic equation (\ref{sys:nodelays}).}
\label{table:p0i}
\end{table}

\begin{table}[h]
	\begin{tabular}{| l| l|}
		\hline
		$p_{10} = an_2D[ad_gM + n_1h(a + d_g)]$  & $p_{13} = -{M}^{2}bb_{{3}} \left( b_{{1}}n_{{3}}+b \right)$ 	\\ 
		\hline

		\multicolumn{2}{|l|}{  $p_{11} = n_2a(bp-b_2d_gT_1)M^2 + a[n_2(ab_3D-n_1hb_2T_1) +b_3d_gG(1-2n_2T_2)]M$} \\
		\hline	
		\multicolumn{2}{|l|}{ $p_{12} = ahb_{{3}}n_{{1}}+ \left[ab_2n_2T_1(b-b_3) -b^2(a + d_g) \right] {M}^{2}-a{b_{{3}}}^{2}n_{{2}}T_2GM $ } \\
		\hline

	\end{tabular}
\caption{Coefficients $p_{1i}$, $i=0,1,2,3$, from the characteristic equation (\ref{sys:nodelays}).}
\label{table:p1i}
\end{table}

\begin{table}[h]
	\begin{tabular}{| l| l|}
		\hline
		$p_{20} = an_2D[n_1h + M(a + d_g)]$  & $p_{22} = n_1hb_3 + M(ab_3-b^2M)$ \\
		\hline
		\multicolumn{2}{|l|}{  $p_{21}  =  -ab_{{2}}n_{{2}}T_1{M}^{2}+ \left[  \left( b_{{3}}d_{{g}}-ab_{{3}}n_{{2}}T_2\right) G+ab_{{3}}n_{{2}}D+ad_{{g}} \right] M+ \left( a+d_{{g}}
			\right) n_{{1}}h$} \\
		\hline	
	\end{tabular}
\caption{Coefficients $p_{2i}$, $i=0,1,2$, from the characteristic equation (\ref{sys:nodelays}).}
\label{table:p2i}
\end{table}

\end{document}